 \documentclass[aps, twocolumn, showpacs, amsmath]{revtex4-2}
\usepackage{dcolumn}
\usepackage{bm}
\usepackage{graphicx}
\usepackage{color}
\usepackage{hyperref}
\usepackage{amsthm}
\usepackage{amssymb}

\begin{document}
\def\be{\begin{equation}}
\def\ee{\end{equation}}

\def\bc{\begin{center}}
\def\ec{\end{center}}
\def\bea{\begin{eqnarray}}
\def\eea{\end{eqnarray}}
\newcommand{\avg}[1]{\langle{#1}\rangle}
\newcommand{\Avg}[1]{\left\langle{#1}\right\rangle}

\def\ie{\textit{i.e.}}
\def\etal{\textit{et al.}}
\def\m{\vec{m}}
\def\G{\mathcal{G}}

\newcommand{\davide}[1]{{\bf\color{blue}#1}}
\newcommand{\gin}[1]{{\bf\color{green}#1}}

\title{Extremal statistics for a resetting Brownian motion before its first-passage time }
\author{Wusong Guo}
\author{Hao Yan}
\author{Hanshuang Chen}\email{chenhshf@ahu.edu.cn}
\affiliation{School of Physics and Optoelectronic Engineering, Anhui University, Hefei 230601, China}
\begin{abstract}
We study the extreme value statistics of a one-dimensional resetting Brownian motion (RBM) till its first passage through the origin starting from the position $x_0$ ($>0$). By deriving the exit probability of RBM in an interval $\left[0, M \right] $ from the origin, we obtain the distribution $P_r(M|x_0)$ of the maximum displacement $M$  and thus gives the expected value $\langle M \rangle$ of $M$ as functions of the resetting rate $r$ and $x_0$. We find that $\langle M \rangle$ decreases monotonically as $r$ increases, and tends to $2 x_0$ as $r \to \infty$. In the opposite limit, $\langle M \rangle$ diverges logarithmically as $r \to 0$. Moreover, we derive the propagator of RBM in the Laplace domain in the presence of both absorbing ends, and then leads to the joint distribution $P_r(M,t_m|x_0)$ of $M$ and the time $t_m$ at which this maximum is achieved in the Lapalce domain by using a path decomposition technique, from which the expected value $\langle t_m \rangle$ of $t_m$ is obtained explicitly. Interestingly, $\langle t_m \rangle$ shows a nonmonotonic dependence on $r$, and attains its minimum at an optimal $r^{*} \approx 2.71691 D/x_0^2$, where $D$ is the diffusion coefficient. Finally, we perform extensive simulations to validate our theoretical results.

\end{abstract}
\maketitle
\section{Introduction}
Understanding extreme events that occur very infrequently is important as they may bring catastrophic consequences. From natural calamities like earthquake, tsunamis and floods to economic collapses and outbreak of pandemic are all examples of extreme events which can lead to devastating consequences \cite{fisher1928limiting,gumbel1958statistics,leadbetter2012extremes,bouchaud1997universality,davison2015statistics,albeverio2006extreme}. 
Extreme-value statistics (EVS) has been a branch of statistics which deals with the extreme deviations of a random process from its mean behavior. Gnedenko’s classical law of extremes is a general result regarding the asymptotic distribution of extreme value for independent and identically distributed random variables \cite{gnedenko1943distribution}. The study of
EVS has been extremely important in the field of disordered
systems \cite{fyodorov2008freezing,fyodorov2010multifractality}, fluctuating interfaces \cite{raychaudhuri2001maximal,majumdar2004exact}, interacting
spin systems \cite{bar2016exact}, stochastic transport models \cite{majumdar2010real,guillet2020extreme}, random matrices \cite{dean2006large,majumdar2009large,majumdar2014top}, epidemic outbreak \cite{dumonteil2013spatial}, binary search trees \cite{krapivsky2000traveling} and related computer search algorithms \cite{majumdar2002extreme,majumdar2003traveling}. In recent years, there is an increasing interest in studying the extreme value for many weakly and strongly correlated stochastic processes \cite{majumdar2010universal,schehr2014exact,lacroix2020universal,PhysRevLett.111.240601,PhysRevLett.117.080601,PhysRevLett.129.094101,PhysRevLett.130.207101}. We refer the readers to \cite{fortin2015applications,majumdar2020extreme} for two recent reviews on the extreme-value statistics.

One of the central goals on this subject is to compute
the statistics of extremes, i.e., the maximum $M$ of a given
trajectory $x(t)$ during an observation time window $\left[0, t \right] $, and the time $t_m$ at which the maximum $M$ is reached. A paradigmatic example is the one-dimensional Brownian motion for a fixed duration $t$ starting from the origin. The joint distribution of $M$ and $t_m$ is given by \cite{schehr2010extreme}
\begin{eqnarray}\label{eq0.1}
P_0(M, t_m|t)=\frac{M}{{2\pi D\sqrt {t_m^3\left( {t - {t_m}} \right)} }}{e^{ - {M^2}/4Dt_m}}
\end{eqnarray}
where $D$ the diffusion coefficient. Integrating $P_0(M,t_m|t)$ over $t_m$ from 0 to $t$, one can get the marginal distribution of $M$,  
\begin{eqnarray}\label{eq0.2}
P_0(M|t)= \frac{1}{\sqrt{\pi D t}} e^{-M^2/4Dt}, \quad M>0, 
\end{eqnarray}
which is the one-sided Gaussian distribution. Integrating $P_0(M,t_m|t)$ over $M$ from 0 to $\infty$, one can obtain the marginal distribution of $t_m$,  
\begin{eqnarray}\label{eq0.3}
P_0(t_m|t)= \frac{1}{\pi \sqrt{t_m(t-t_m)}}, \quad 0\leq t_m \leq t,
\end{eqnarray}
which is often referred to as the ``arcsine law" due to P. L\'evy \cite{Levy1940ArcsineLaw,feller1971introduction,majumdar2007brownian}. The name stems from the fact that the cumulative distribution of $t_m$ reads $F(z) =  \int_0^z {P(t_m )dt_m}  = (2/\pi)\arcsin \sqrt {z/t}$. A counterintuitive aspect of the $U$-shaped distribution Eq.(\ref{eq0.3}) is that its average value $\langle t_m \rangle=t/2$ corresponds to the minimum of the
distribution, i.e., the less probable outcome, whereas values
close to the extrema $t_m=0$ and $t_m=t$ are much more likely.
Recent studies led to many extensions of the law, such as in constrained Brownian motions \cite{majumdar2008time}, random acceleration process \cite{majumdar2010time,boutcheng2016occupation}, fractional Brownian motion \cite{sadhu2018generalized,sadhu2021functionals}, run-and-tumble motion \cite{SinghArcsinelaws_RTP}, resetting Brownian motion \cite{PhysRevE.103.052119}, and for general
stochastic processes \cite{lamperti1958occupation,kasahara1977limit,dhar1999residence,majumdar2002exact,schehr2010extreme,PhysRevLett.107.170601,PhysRevE.83.061146,PhysRevE.105.024113,PhysRevE.102.032103}. Extension to study the distribution of the time difference
between the minimum and the maximum for stochastic processes has also been made in \cite{mori2019time,mori2020distribution}. Quite remarkably, the statistics of $t_m$ has found applications in convex hull problems \cite{randon2009convex} and also in detecting whether a stationary process is equilibrium or not \cite{mori2021distribution,mori2022time}.

While the statistics of $M$ and $t_m$  in a fixed duration time has been extensively studied, the study of these quantities for a stochastic process until a stopping time, e.g. the first passage time when the process arrives at some threshold value brings some recent attention. This problem is relevant to some context. For instance, in queue theory the maximum queue length and the time at which this length is achieved before the queue length gets to zero \cite{randon2007distribution}. In stock market, an agent can hold the stock till its price reaches a certain threshold value. A best time to sell the stock is when the price of the stock reaches its maximum before dropping to the threshold \cite{majumdar2008optimal}. Another example arises in the biological context
regarding the maximal excursions of the tracer proteins before binding at a site \cite{RevModPhys.83.81,PhysRevX.7.011019}. For a one-dimensional Brownian motion starting from the position $x_0$($>0$), the statistics of $M$ and $t_m$ before its first passage through the origin has been studied, and the marginal distributions of $M$ and $t_m$ are given by \cite{randon2007distribution}
\begin{eqnarray}\label{eq0.4}
P_0(M|x_0)= \frac{x_0}{M^2}, \quad M \geq x_0,
\end{eqnarray}
and 
\begin{eqnarray}\label{eq0.5}
P_0( {{t_m}|{x_0}} ) = \frac{1}{{2\pi {t_m}}}\left[ {\pi  - \int_0^\pi  {dy\vartheta_4 \left( {y/2,{e^{ - D{t_m}{y^2}/x_0^2}}} \right)} } \right]
\end{eqnarray}
where $\vartheta_4(y,z)$ is the fourth of Jacobi's theta functions. It was shown that $P_0(t_m|x_0)$ exhibits power-law forms at both large and small tails with $P_0(t_m|x_0) \sim t_m^{-1/2}$ as $t_m \to 0$ and $P_0(t_m|x_0) \sim t_m^{-3/2}$ as $t_m \to \infty$. A recent study has extended to compute the joint distribution of $M$ and $t_m$ and their marginal distributions for the run-and-tumble particle in one dimension \cite{singh2022extremal}.

Stochastic resetting is a renewal process in which the
dynamics is interrupted stochastically followed by its starting anew. The subject has recently gained considerable
attention due to wide applications in search problems \cite{PhysRevLett.113.220602,PhysRevE.92.052127}, the optimization of randomized computer algorithms \cite{PhysRevLett.88.178701}, and in the field of
biophysics \cite{reuveni2014role,rotbart2015michaelis} (see \cite{evans2020stochastic,Gupta2022Review} for two recent reviews). A paradigmatic example in statistical physics is resetting Brownian motions where a diffusing particle is reset to its starting point at random times but with a constant rate. A finite resetting rate leads to a nonequilibrium stationary state with non-Gaussian fluctuations for the
particle position. The mean time to reach a given target for the first time can become finite and be minimized with respect to the resetting rate \cite{evans2011diffusion}. Some extensions have been made in the field, such as spatially \cite{evans2011diffusion2} or temporally \cite{NJP2016.18.033006,pal2016diffusion,PhysRevE.93.060102,PhysRevE.96.012126,PhysRevE.100.032110} dependent resetting rate,  higher dimensions \cite{Evans2014_Reset_Highd}, complex geometries \cite{Christou2015,PhysRevResearch.2.033027,BressloffJSTAT2021,PhysRevE.105.034109}, noninstantaneous resetting \cite{EvansJPA2018,PalNJP2019,PhysRevE.101.052130,GuptaJPA2020}, in the presence of external potential \cite{pal2015diffusion,ahmad2019first,gupta2020stochastic}, other types of Brownian motion, like run-and-tumble particles \cite{evans2018run,santra2020run,bressloff2020occupation}, active particles \cite{scacchi2018mean,kumar2020active}, constrained Brownian particle \cite{PhysRevLett.128.200603}, and so on \cite{basu2019symmetric}.  
These nontrivial findings have triggered an enormous recent activities in the field, including statistical physics \cite{PhysRevLett.116.170601,pal2017first,gupta2014fluctuating,evans2014diffusion,meylahn2015large,chechkin2018random,PhysRevE.103.022135,de2020optimization,magoni2020ising,arXiv:2202.04906,JPA2022.55.021001,JPA2022.55.234001,sokolov2023linear}, stochastic thermodynamics \cite{fuchs2016stochastic,pal2017integral,gupta2020work}, chemical and biological
processes \cite{reuveni2014role,PhysRevLett.112.240601,rotbart2015michaelis,PhysRevLett.128.148301,JPA2022.55.274005}, record statistics \cite{JStatMech2022.063202,JPA2022.55.034002,kumar2023universal,smith2023striking}, optimal control theory \cite{arXiv:2112.11416}, and single-particle experiments \cite{tal2020experimental,besga2020optimal}.

In the present work, we aim to study the statistics of $M$ and $t_m$ for a resetting Brownian particle in one dimension till it passes through the origin for the first time, starting from a positive position $x_0$. We first compute analytically the marginal distribution $P_r(M|x_0)$ of $M$ by the splitting or exit probability from the origin when the resetting Brownian motion is confined in an interval $\left[0, M \right]$. For any nonzero resetting rate $r$, $P_r(M|x_0)$ decays exponentially in the large-$M$ limit, such that the expectation $\langle M \rangle$ of $M$ converges. The exact expression of $\langle M \rangle$ is also derived as functions of $r$ and $x_0$. $\langle M \rangle$ diverges with $-\frac{x_0}{2}\ln r$ as $r \to 0$, and decreases monotonically with $r$ and converges to $2 x_0$ as $r \to \infty$. Using the path decomposition technique, we express the joint distribution $P_r(M,t_m|x_0)$ of $M$ and $t_m$ in the Laplace space, from which we can obtain the exact expression of the expectation $\langle t_m \rangle$ of $t_m$. Interestingly, $\langle t_m \rangle$ shows a unique minimum at an optimal resetting rate $r^{*}\approx 2.71691 D/x_0^2$, reminiscent of another optimal resetting rate $\approx 2.53964 D/x_0^2$ at which the mean first passage time is a minimum \cite{evans2011diffusion,evans2020stochastic}.

\begin{figure}
	\centerline{\includegraphics*[width=1.0\columnwidth]{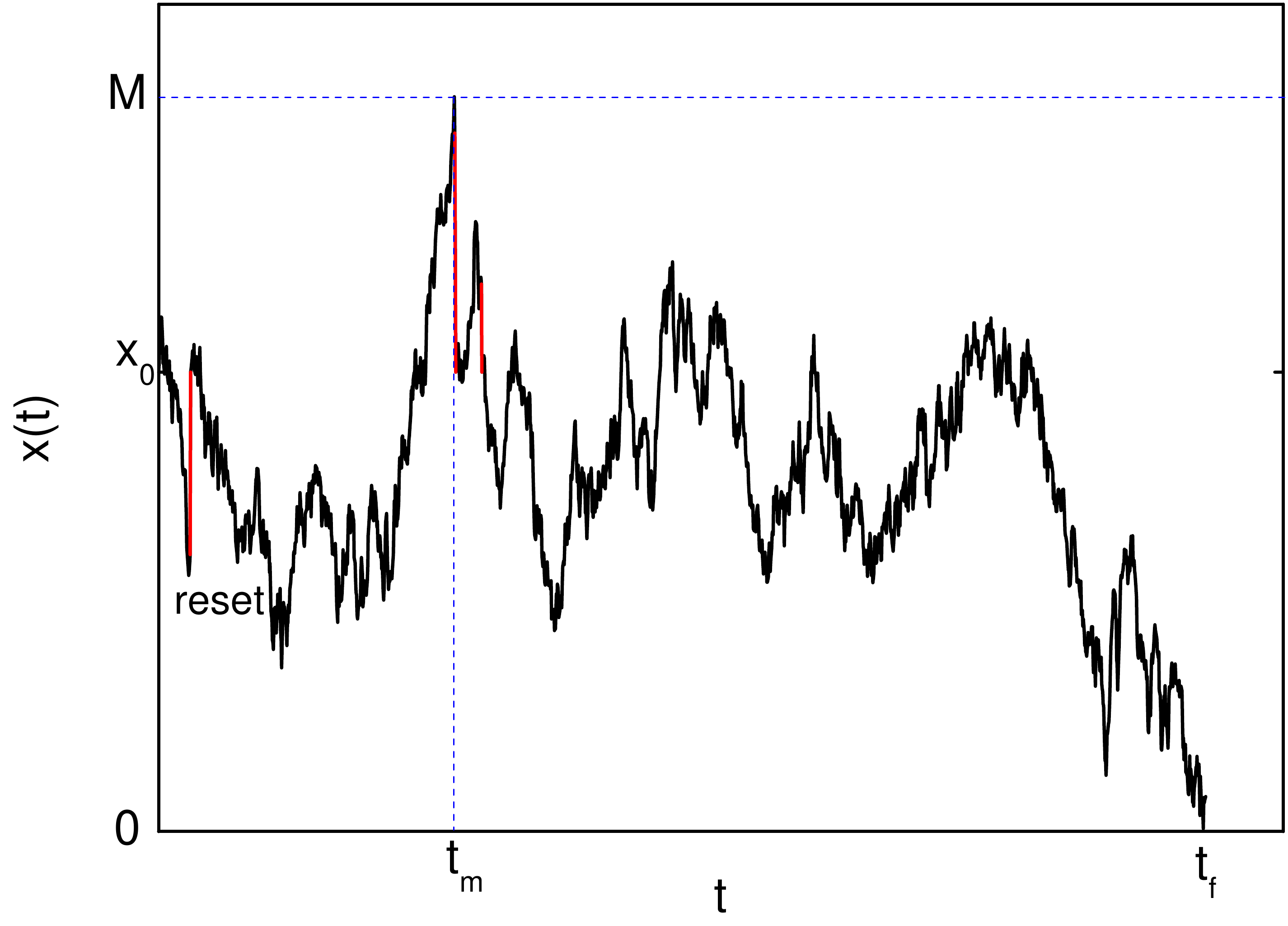}}
	\caption{A realization of a one-dimensional resetting Brownian motion in the presence of an absorbing wall at the origin. The stochastic process $x(t)$ starts from $x_0$ and reaches its maximum $M$ at time $t_m$ before the first-passage time $t_f$ through the origin. \label{fig1}}
\end{figure}

\section{Model}
Let us consider a one-dimensional resetting Brownian motion (RBM), starting at $x_0$($>0$), with resetting to the position $x_r>0$ at rate $r$. The position $x(t)$ of the particle at time $t$ is updated by the following stochastic rule \cite{evans2011diffusion,evans2020stochastic}:
\begin{eqnarray}\label{eq1.0}
{x}( {t + dt} ) = \left\{ \begin{array}{llc}
{x}( t ) + \sqrt {2 D } {\xi }(t) dt , &  {\rm{with}} \, {\rm{prob.}}  & 1-r dt,  \\
{x}_r, &   {\rm{with}} \, {\rm{prob.}}   & r dt,  \\ 
\end{array}  \right. 
\end{eqnarray}
where $D$ is the diffusion coefficient, and ${\xi }(t)$ is a Gaussian white noise with zero mean $\langle {\xi (t)} \rangle  = 0$ and delta correlator $\langle {\xi (t)}{\xi (t'} \rangle  = \delta ( {t - t'} )$.

We assume that there is an absorbing boundary located at the origin $x=0$, such that the stochastic process is terminated once the particle hits the boundary. In Fig.\ref{fig1}, we show a realization of the RBM $x(t)$, starting at $x_0$ ($>0$) and terminating whenever $x(t)$ reaches the origin. The displacement $x(t)$ reaches its maximum $M$ at time $t_m$. We are interested in the joint distribution $P_r(M, t_m|x_0)$ of the maximum displacement $M$ of the particle and the time $t_m$ at which the maximum is reached before the first passage time to the absorbing boundary at the origin, and their marginal distributions of $M$ and $t_m$, $P_r(M|x_0)$ and $P_r(t_m|x_0)$.

\section{Exit probability}
Consider the RBM starting from $x_0\in \left[ 0,M\right] $, and both of the interval are absorbing boundaries. Let us denote by $\mathcal{E}_r(x_0)$ the splitting or exit probability that the particle exits the interval for the first time through the origin, i.e., the probability that the maximum before the first-passage time is less than or equal to $M$. The exit probability has been obtained in a recent work \cite{PhysRevE.99.032123}. For completeness, we derive the expression but use a different method. $\mathcal{E}_r(x_0)$ satisfies the following equation \cite{redner2001guide}, 
\begin{eqnarray}\label{eq1.1}
D\frac{{{d^2}\mathcal{E}_r\left( {{x_0}} \right)}}{{dx_0^2}} - r\mathcal{E}_r\left( {{x_0}} \right) + r\mathcal{E}_r\left( {{x_r}} \right) = 0,
\end{eqnarray}
Eq.(\ref{eq1.1}) can be solved subject to the boundary conditions
\begin{eqnarray}\label{eq1.2}
\mathcal{E}_r(0)=1,\quad  \mathcal{E}_r(M)=0,
\end{eqnarray}
which yields
\begin{eqnarray}\label{eq1.3}
\mathcal{E}_r\left( {{x_0}} \right) = 
\frac{{\sinh \left[ {{\alpha _0}\left( {M - {x_r}} \right)} \right] + \sinh \left[ {{\alpha _0}\left( {{x_r} - {x_0}} \right)} \right]}}{{\sinh \left[ {{\alpha _0}\left( {M - {x_r}} \right)} \right] + \sinh \left( {{\alpha _0}{x_r}} \right)}},
\end{eqnarray}
where $\alpha _0=\sqrt{r/D}$. For $x_r=x_0$, Eq.(\ref{eq1.3}) simplies to 
\begin{eqnarray}\label{eq1.4}
\mathcal{E}_r\left( {{x_0}} \right) =\frac{{\sinh \left[ {{\alpha _0}\left( {M - {x_0}} \right)} \right]}}{{\sinh \left[ {{\alpha _0}\left( {M - {x_0}} \right)} \right] + \sinh \left( {{\alpha _0}{x_0}} \right)}}.
\end{eqnarray}

\section{Marginal distribution of the maximum displacement before the first passage to the origin}
Differentiating Eq.(\ref{eq1.4}) with respect to $M$ gives the probability density function of $M$ \cite{randon2007distribution},
\begin{eqnarray}\label{eq1.5}
P_r\left( M|x_0 \right) &=& \frac{\partial \mathcal{E}_r}{\partial M} \nonumber \\ &=&\frac{{{\alpha _0}\sinh \left[ {{\alpha _0}{x_0}} \right]\cosh\left[ {{\alpha _0}\left( {M - {x_0}} \right)} \right]}}{{{{\left( {\sinh \left[ {{\alpha _0}\left( {M - {x_0}} \right)} \right] + \sinh \left[ {{\alpha _0}{x_0}} \right]} \right)}^2}}} .
\end{eqnarray}
In the limit $r \to 0$, Eq.(\ref{eq1.5}) recovers to the result when the resetting is absent, see Eq.(\ref{eq0.4}).

\begin{figure}
	\centerline{\includegraphics*[width=1.0\columnwidth]{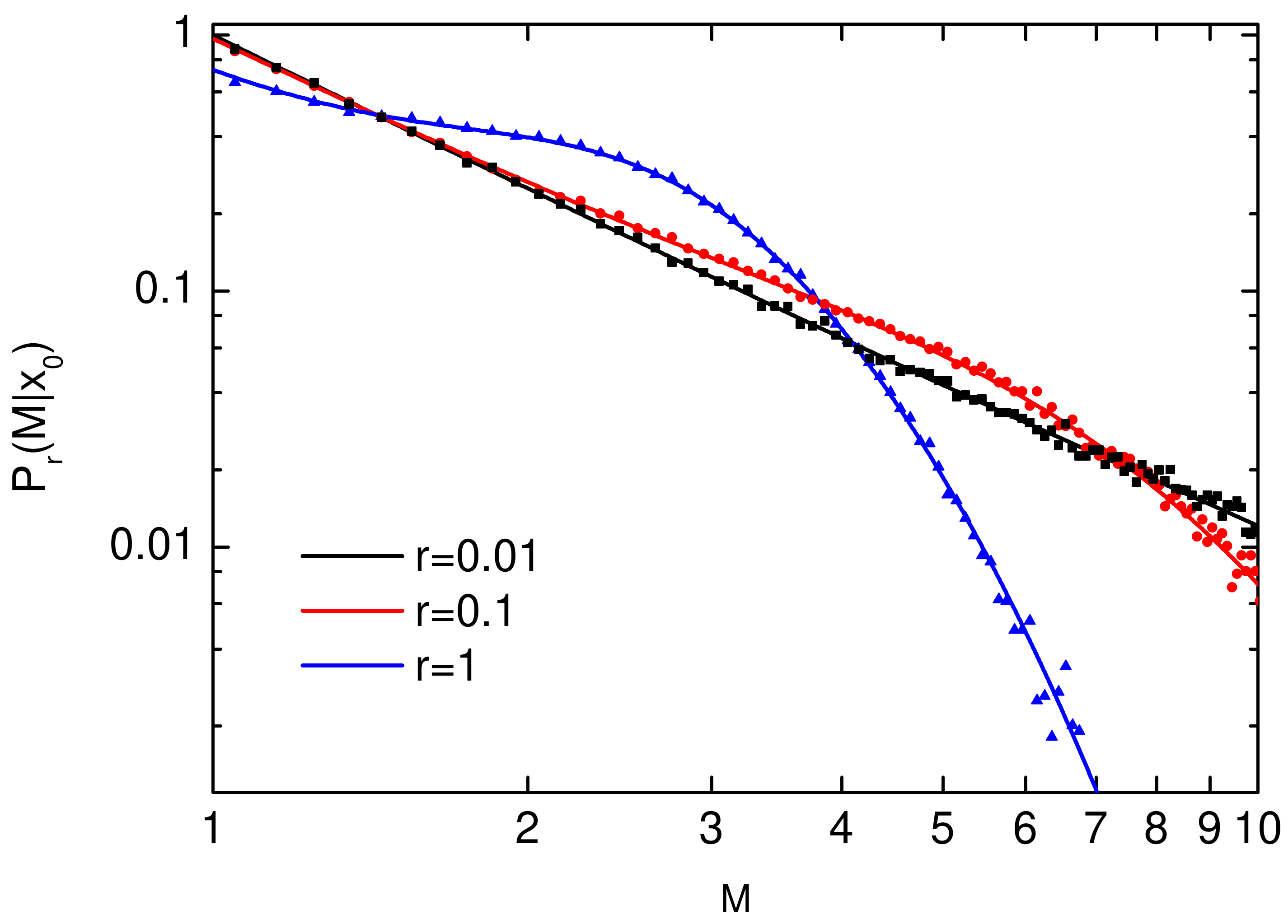}}
	\caption{The marginal distribution $P_r(M|x_0)$ of the maximum displacement $M$ for three different resetting rate $r$, where  $x_0=1$ and $D=1/2$ are fixed. The lines and symbols correspond to the theoretical and simulation results, respectively. \label{fig2}}
\end{figure}

In Fig.\ref{fig2}, we plot the distribution $P_r(M|x_0)$ for three different values of $r$ but for the fixed $x_0=1$ and $D=1/2$. For small resetting rates, $P_r(M|x_0)$ decays with $M$ very slowly like a power law, similar to the case without resetting (see Eq.(\ref{eq0.4})). For relatively larger resetting rates, the resetting to the starting position can decrease the long-range meandering of diffusing particle, so that the fat-tailed distribution is cut off in the large-$M$ limit.      

The expectation of $M$ can be computed as
\begin{eqnarray}\label{eq1.7}
\langle M(x_0) \rangle =\int_{x_0}^{\infty} dM M P_r(M|x_0)=\alpha _0^{-1}F_1(\gamma ),
\end{eqnarray}
where
\begin{eqnarray}\label{eq1.7.2}
\gamma=x_0 \alpha_0 =x_0 \sqrt{r/D},
\end{eqnarray}
and
\begin{eqnarray}\label{eq1.7.1}
F_1(\gamma ) =\gamma + \tanh(\gamma) \left[ {\gamma + \ln \left( {\coth \left( {\gamma/2} \right)} \right)} \right].
\end{eqnarray}
For $r=0$, $\langle M(x_0) \rangle$ is divergent. This is because that the standard Brownian motion can diffuse indefinitely in the positive direction without touching the absorbing boundary at the origin. For any nonzero $r$, $P_r(M|x_0)$ decays exponentially with $M$ in the large-$M$ limit, $P_r(M|x_0) \sim e^{-\alpha_0 M}$, and thus $\langle M(x_0) \rangle$ becomes convergent. In the limit of $r \to 0$,
\begin{eqnarray}\label{eq1.8}
\langle M(x_0) \rangle   \sim -\frac{x_0}{2} \ln r, \quad  r \to 0.
\end{eqnarray}
Therefore, the expected maximum $\langle M(x_0) \rangle$ diverges logarithmically as $r\to 0$.  In the opposite limit $r \to \infty$, 
\begin{eqnarray}\label{eq1.9}
\langle M(x_0) \rangle =  2 x_0, \quad  r \to \infty.
\end{eqnarray}
This result is a bit surprising as one may expect intuitively that the expected maximum is $x_0$ rather than $2 x_0$ as $r \to \infty$. 

\begin{figure}
	\centerline{\includegraphics*[width=1.0\columnwidth]{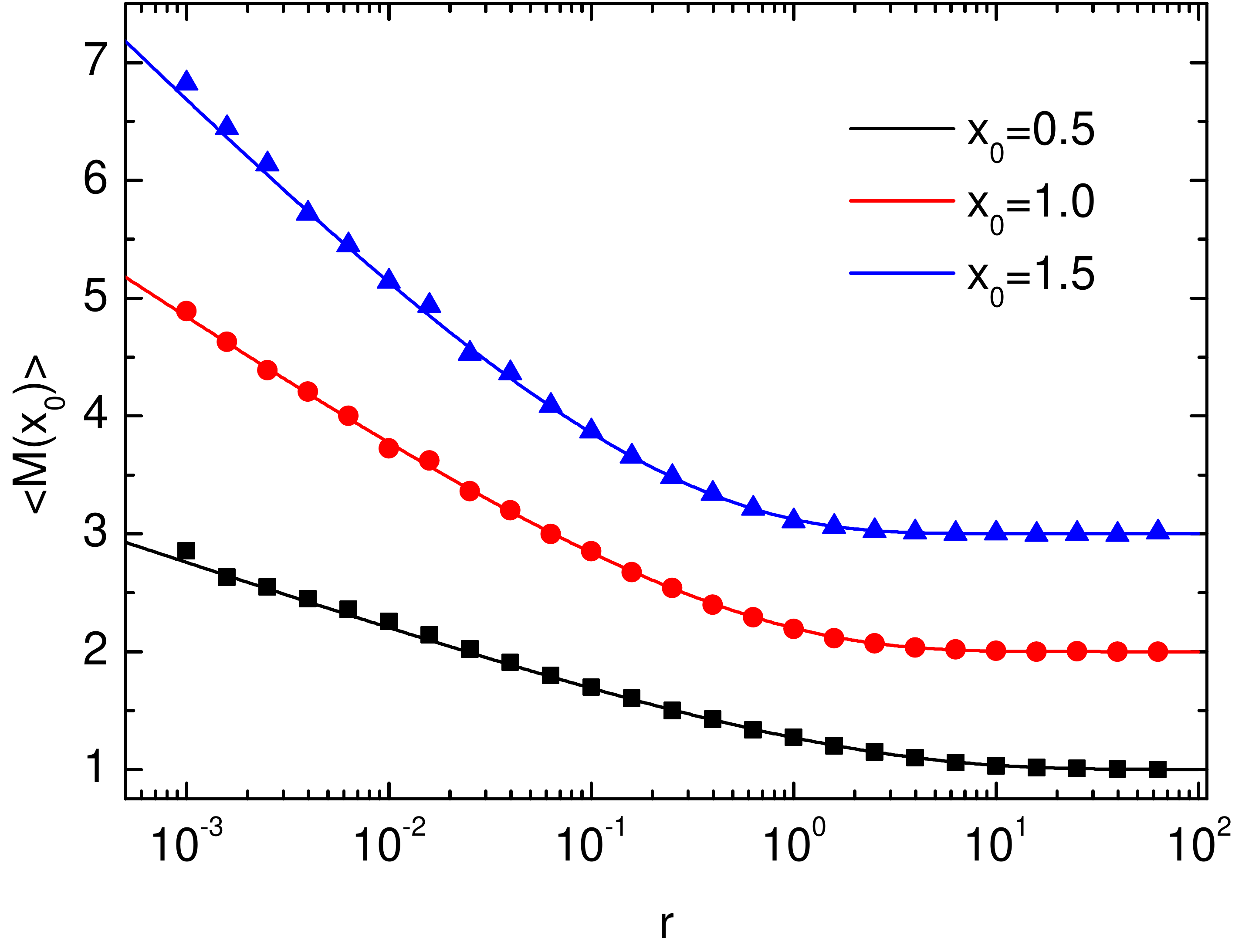}}
	\caption{The expected value $\langle M(x_0) \rangle$ of the maximum displacement $M$ as a function the resetting rate $r$ for three different values of $x_0$, where $D=1/2$ is fixed. The lines and symbols correspond to the theoretical and simulation results, respectively. \label{fig3}}
\end{figure}

In Fig.\ref{fig3}, we show $\langle M(x_0) \rangle$ as a function of $r$ for three different values of $x_0$. Clearly, $\langle M(x_0) \rangle$ decreases monotonically with $r$ and approaches to $2 x_0$ in the limit of $r \to \infty$. Simulation results are also shown in Fig.\ref{fig3} and support the theoretical predictions. 

\section{Survival probability}
In this section, we compute the survival probability $Q_r(x_0,t)$ of the RBM confined in an interval $\left[ 0,M\right]$ with the absorbing boundaries at both ends, defined as the probability that the particle has hit neither of the boundaries until time $t$ starting from $x_0 \in \left( 0,M \right)$. The backward equation for $Q_r(x_0,t)$ reads \cite{evans2011diffusion,evans2011diffusion2,evans2020stochastic}
\begin{eqnarray}\label{eq2.1}
\frac{{\partial {Q_r}\left( {{x_0},t} \right)}}{{\partial t}} = D\frac{{{\partial ^2}{Q_r}\left( {{x_0},t} \right)}}{{\partial x_0^2}} - r{Q_r}\left( {{x_0},t} \right) + r{Q_r}\left( {{x_r},t} \right),\nonumber \\
\end{eqnarray}
where the boundary conditions are $Q_r(0,t)=Q_r(M,t)= 0$ and the initial condition is $Q_r(x_0,0)=1$.

Performing the Laplace transform $\tilde{Q}_r(x_0,s)=\int_{0}^{\infty} e^{-st} Q_r(x_0,t)dt$, Eq.(\ref{eq2.1}) becomes
\begin{eqnarray}\label{eq2.2}
D\frac{{{d^2}{{\tilde Q}_r}\left( {{x_0},s} \right)}}{{dx_0^2}} - (r + s){{\tilde Q}_r}\left( {{x_0},s} \right) + r{{\tilde Q}_r}\left( {{x_r},s} \right) =  - 1,\nonumber \\
\end{eqnarray}
subject to the boundary conditions $\tilde{Q}_r(0,s)=\tilde{Q}_r(M,s)= 0$. The solution of Eq.(\ref{eq2.2}) reads
\begin{eqnarray}\label{eq2.3}
{{\tilde Q}_r}( {{x_0},s} ) =\frac{{\sinh ( {\alpha M} ) - \sinh ( {\alpha {x_0}} ) - \sinh \left[ {\alpha ( {M - {x_0}} )} \right]}}{{s\sinh ( {\alpha M} ) + r\sinh ( {\alpha {x_r}} ) + r\sinh \left[ {\alpha ( {M - {x_r}} )} \right]}},\nonumber \\
\end{eqnarray}
where $\alpha=\sqrt{(r+s)/D}$. For $x_r=x_0$, Eq.(\ref{eq2.3}) simplies to
\begin{eqnarray}\label{eq2.4}
{{\tilde Q}_r}( {{x_0},s} ) =\frac{{\sinh ( {\alpha M} ) - \sinh ( {\alpha {x_0}} ) - \sinh \left[ {\alpha ( {M - {x_0}} )} \right]}}{{s\sinh \left( {\alpha M} \right) + r\sinh ( {\alpha {x_0}} ) + r\sinh \left[ {\alpha ( {M - {x_0}} )} \right]}}.\nonumber \\
\end{eqnarray}
Eq.(\ref{eq2.4}) was also obtained in a recent work \cite{PhysRevE.99.032123}. 

\section{Propagator}
Let us denote by $G_r(x,t|x_0)$ the propagator of the RBM in an interval $\left[ 0,M\right] $ with the absorbing boundaries at both ends. One
can write a time-dependent equation for the propagator
$G_r(x,t|x_0)$ using a last renewal formalism  \cite{evans2011diffusion,evans2011diffusion2,evans2020stochastic}
\begin{eqnarray}\label{eq3.1}
{G_r}\left( {x,t|{x_0}} \right) &=& {e^{ - rt}}{G_0}\left( {x,t|{x_0}} \right) \nonumber \\ &+& r\int_0^t {d\tau {e^{ - r\tau }}{G_0}\left( {x,\tau |{x_0}} \right){Q_r}\left( {{x_0},t - \tau } \right)}. \nonumber \\
\end{eqnarray}
Here, $G_0(x,t|x_0)$ is the propagator in the absence of resetting. Also, recall that $Q_r(x_0,t)=\int_{0}^{M} dx G_r(x,t|x_0)$ is the survival probability until time $t$. The first term on the right-hand side of Eq.(\ref{eq3.1}) corresponds to the case when no resetting happens in the interval $\left[0, t \right] $ with the probability $e^{rt}$. The second term refers to the case when at least one resetting happens in the interval $\left[0, t \right]$, where the last resetting occurs at $t-\tau$ with the probability $r e^{-r \tau} d \tau$. The factor $Q_r(x_0, t-\tau)$ is the survival probability during
this interval $\left[0, t-\tau \right]$.

It is convenient to take the Laplace transform of Eq.(\ref{eq3.1}) so that the convolution structure in the second term on the right-hand side of Eq.(\ref{eq3.1}) can be exploited. By defining $\tilde{G}_r(x,s|x_0)=\int_{0}^{\infty} e^{-st}G(x,t|x_0)dt$, Eq.(\ref{eq3.1}) becomes
\begin{eqnarray}\label{eq3.2}
{{\tilde G}_r}\left( {x,s|{x_0}} \right) = {{\tilde G}_0}\left( {x,r + s|{x_0}} \right)\left[ {1 + r{{\tilde Q}_r}\left( {{x_0},s} \right)} \right].
\end{eqnarray}
Here, the Laplace transform $\tilde G_0(x,s|x_0)$ of the propagator $G_0(x,t|x_0)$ without resetting a classical result and known from the literature \cite{redner2001guide},
\begin{widetext}
\begin{eqnarray}\label{eq3.3}
{{\tilde G}_0}\left( {x,s|{x_0}} \right) = \frac{{\cosh \left[ {\sqrt {\frac{s}{D}} \left( {M - \left| {x - {x_0}} \right|} \right)} \right] - \cosh \left[ {\sqrt {\frac{s}{D}} \left( {M - x - {x_0}} \right)} \right]}}{{2\sqrt {sD} \sinh \left( {\sqrt {\frac{s}{D}} M} \right)}}.
\end{eqnarray}
Substituting Eq.(\ref{eq2.4}) and Eq.(\ref{eq3.3}) into Eq.(\ref{eq3.2}), we obtain
\begin{eqnarray}\label{eq3.4}
{\tilde G}_r(x,s|{x_0}) = \frac{\alpha}{2}\frac{\cosh \left[\alpha (M-|x-x_0|) \right] -\cosh \left[\alpha (M-x-x_0) \right]  }{s \sinh(\alpha M)+r \sinh(\alpha x_0)+r \sinh\left[\alpha (M-x)) \right] },
\end{eqnarray}
where $\alpha=\sqrt{(r+s)/D}$ again. Eq.(\ref{eq3.4}) was also obtained in a recent work \cite{PhysRevE.99.032123}. 
\end{widetext}

\section{Joint distribution of $M$ and $t_m$ before its first passage to the origin}
Let us define $P_r(M,t_m|x_0)$ as the joint probability density function that the RBM reaches its maximum $M$ at time $t_m$ before passing through the origin for the first time $t_f$, providing that the Brownian starts from the position $x_0$ ($>0$). To compute the joint distribution $P_r(M,t_m|x_0)$, we can decompose the trajectory into two parts: a left-hand segment for which $0<t<t_m$, and a right-hand segment for which $t_m<t<t_f$. The statistical weight of the first segment equals to the propagator $G_r(M,t_m|x_0)$. However, it turns out that $G_r(M,t_m|x_0)=0$ which implies that the contribution from this part is zero. To circumvent this problem, we compute $G_r(M-\epsilon,t_m|x_0)$ and later take the limit $\epsilon \to 0$ \cite{majumdar2004exact}. The statistical weight of the second segment is given by the exit probability $\mathcal{E}_r(M-\epsilon)$. Due to the Markov property, the joint probabilty density $P_r(M,t_m|x_0)$ can be written as the product of the statistical weights of two segments \cite{randon2007distribution}, 
\begin{eqnarray}\label{eq4.1}
{P_r}\left( {M,{t_m}|{x_0}} \right) = \lim_{\epsilon \to 0} \mathcal{N}( {{x_0},\epsilon} ){G_r}\left( {M - \epsilon,{t_m}|{x_0}} \right){\mathcal{E}_r}\left( {M - \epsilon} \right), \nonumber \\
\end{eqnarray}
where the normalization factor $\mathcal{N}( {{x_0},\epsilon} )$  will be determined later. 

One can compute the Laplace transform of $G_r(M-\epsilon,t_m|x_0)$ in terms of Eq.(\ref{eq3.4}) and ${\mathcal{E}_r}\left( {M - \epsilon} \right) $ by Eq.(\ref{eq1.3}), and keeps the leading order in $\epsilon$, given by
\begin{widetext} 
\begin{eqnarray}\label{eq4.2}
{{\tilde G}_r}\left( {M - \epsilon,s|{x_0}} \right) = \frac{{{\alpha ^2}\sinh \left( {\alpha {x_0}} \right)}}{{s\sinh \left( {\alpha M} \right) + r\sinh \left( {\alpha {x_0}} \right) + r\sinh \left[ {\alpha (M - {x_0})} \right]}}\epsilon,
\end{eqnarray}
and
\begin{eqnarray}\label{eq4.3}
{\mathcal{E}_r}\left( {M - \epsilon} \right) = \frac{{{\alpha _0}\cosh \left[ {{\alpha _0}\left( {M - {x_0}} \right)} \right]}}{{\sinh \left[ {{\alpha _0}\left( {M - {x_0}} \right)} \right] + \sinh \left( {{\alpha _0}{x_0}} \right)}}\epsilon .
\end{eqnarray}

It is conveninent to perform the Laplace transform for $P_r(M,t_m|x_0)$ over $t_m$, 
\begin{eqnarray}\label{eq4.3.1}
\tilde{P}_r(M,s|x_0)&=&\int_{0}^{\infty} d t_m e^{-s t_m} P_r(M,t_m|x_0) \nonumber \\&=& \mathcal{N}( {{x_0},\epsilon} ){\tilde{G}_r}\left( {M - \epsilon,{s}|{x_0}} \right){\mathcal{E}_r}\left( {M - \epsilon} \right). \nonumber \\
\end{eqnarray}
Letting $s \to 0$, the left hand side of Eq.(\ref{eq4.3.1}) is just the marginal distribution $P_r(M|x_0)$, which yields
\begin{eqnarray}\label{eq4.4}
{P_r}\left( {M|{x_0}} \right) &=& \int_0^\infty d{t_m}{P_r}\left( {M,{t_m}|{x_0}} \right) \nonumber \\ &=&  \mathcal{N}\left( {{x_0},\epsilon} \right){{\tilde G}_r}\left( {M - \epsilon,0|{x_0}} \right){\mathcal{E}_r}\left( {M - \epsilon} \right).
\end{eqnarray}
Substituting Eq.(\ref{eq1.5}), Eq.(\ref{eq4.2}) and Eq.(\ref{eq4.3}) into Eq.(\ref{eq4.4}), we obtain
\begin{eqnarray}\label{eq4.5}
\mathcal{N}( {{x_0},\epsilon} ) =\frac{D}{\epsilon^2}.
\end{eqnarray}
which is independent of the starting position $x_0$. Substituting  Eq.(\ref{eq4.2}), Eq.(\ref{eq4.3}) and Eq.(\ref{eq4.5}) into Eq.(\ref{eq4.3.1}), we obtain the joint distribution $P_r(M,t_m|x_0)$ in the Laplace space, 
\begin{eqnarray}\label{eq4.5.1}
\tilde{P}_r(M,s|x_0)=\frac{{\left( {r + s} \right)\sinh \left( {\alpha {x_0}} \right)}}{{s\sinh \left( {\alpha M} \right) + r\sinh \left( {\alpha {x_0}} \right) + r\sinh \left[ {\alpha \left( {M - {x_0}} \right)} \right]}}\frac{{{\alpha _0}\cosh \left[ {{\alpha _0}\left( {M - {x_0}} \right)} \right]}}{{\sinh \left[ {{\alpha _0}\left( {M - {x_0}} \right)} \right] + \sinh \left( {{\alpha _0}{x_0}} \right)}} .
\end{eqnarray}
\end{widetext} 
To obtain the joint distribution $P_r(M,t_m|x_0)$, one has to perform the inverse Laplace transformation for Eq.(\ref{eq4.5.1}) with respect to $s$. Unfortunately, it turns out to be a challenging task. However, one may expect to obtain explicitly the statistics of $t_m$, such as the expectation value of $t_m$. To the end, by integrating Eq.(\ref{eq4.3.1}) over $M$ from $x_0$ to $\infty$, one obtain the Laplace transform of the marginal distribution $P_r(t_m|x_0)$,   
\begin{eqnarray}\label{eq4.6}
{{\tilde P}_r}\left( {s|{x_0}} \right) &=& \int_0^\infty  {d{t_m}} {e^{ - s{t_m}}}{P_r}\left( {{t_m}|{x_0}} \right) \nonumber \\ &=& \int_{x_0}^{\infty} dM \tilde{P}_r(M,s|x_0).
\end{eqnarray}

\begin{figure}
	\centerline{\includegraphics*[width=1.0\columnwidth]{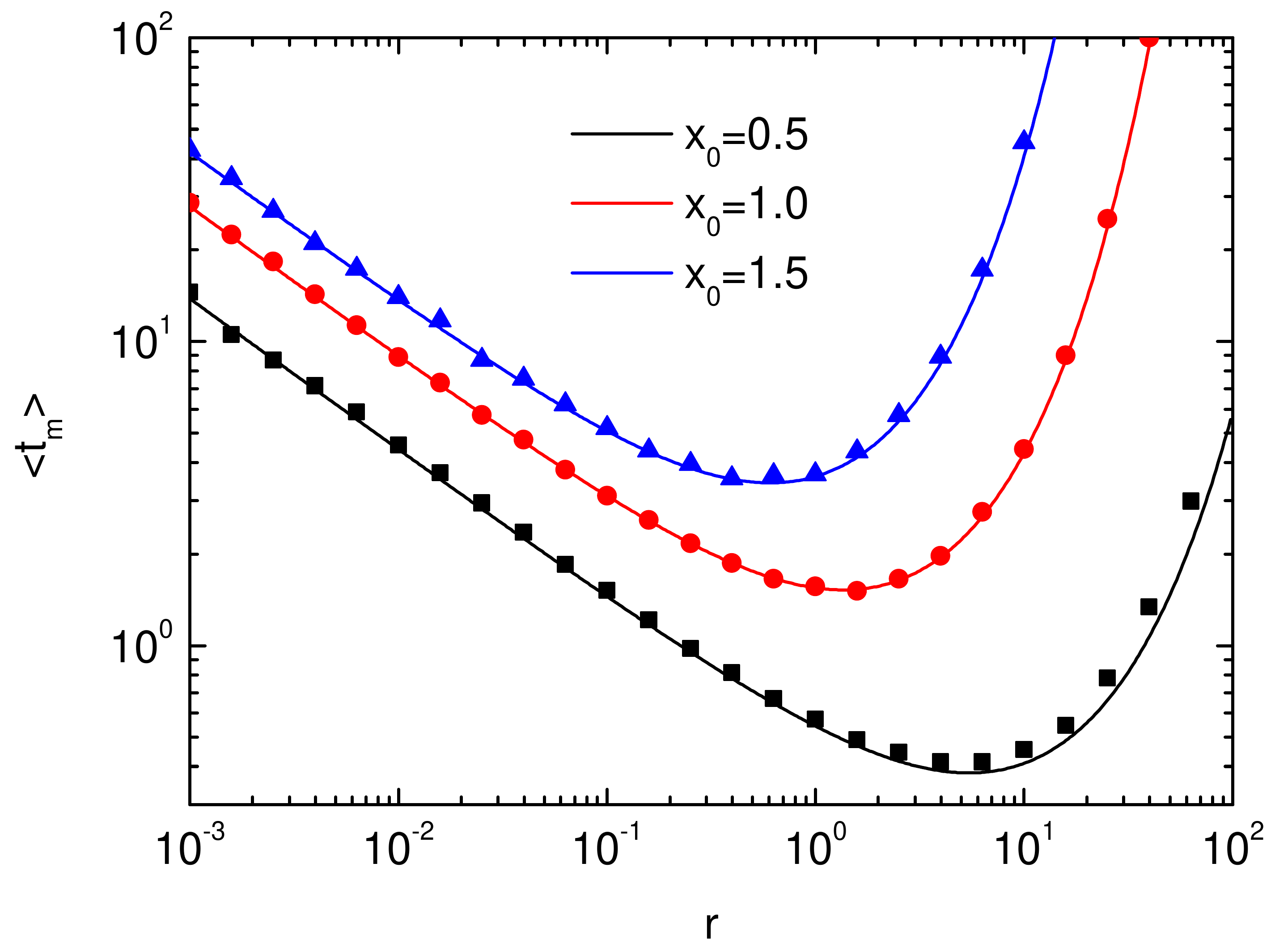}}
	\caption{The expected value $\langle t_m \rangle$ of the time $t_m$ at which the RBM reaches its maximum before its first passage through the origin, having starting from $x_0>0$, as a function of the resetting rate $r$, where  $D=1/2$ is fixed. The lines and symbols correspond to the theoretical and simulation results, respectively. \label{fig4}}
\end{figure}

In particular, the expectation  of the time $t_m$ is given by
\begin{eqnarray}\label{eq4.7}
\langle t_m \rangle=-\lim_{s \to 0} \frac{\partial \tilde{P}_r(s|x_0)}{\partial s}=r^{-1} F_2(\gamma),
\end{eqnarray}
where $\gamma=x_0 \sqrt{r/D}$ and 
\begin{widetext} 
\begin{eqnarray}\label{eq4.6.1}
{F_2}\left( \gamma  \right) &=&  \left\{ {  \coth \left( \gamma  \right)\left[ {  1 + 2{e^{2\gamma }} + {e^{4\gamma }} - \gamma  + 2{e^{2\gamma }} - \gamma {e^{4\gamma }}+ {{\left( {{e^{2\gamma }} - 1} \right)}^2}\ln \left( {\left( {{e^\gamma } + 1} \right)\left( {{e^\gamma } - 1} \right)} \right)} \right] + 4{e^{2\gamma }}\left[ {{\rm{Li}}_2\left( {{e^{ - \gamma }}} \right) - {\rm{Li}}_2\left( { - {e^\gamma }} \right)} \right]} \right\} \nonumber  \\ &\times& \frac{{{e^\gamma }\sinh \left( \gamma  \right)}}{{2{{\left( {1 + {e^{2\gamma }}} \right)}^3}}}  -\frac{\gamma }{4}\coth \left( \gamma  \right) + \frac{{\left( {{e^\gamma } - 1} \right){\rm{csch}} \left( {2\gamma } \right){{\sinh }^2}\left( \gamma  \right)}}{{\left( {1 + {e^\gamma }} \right){{\left( {1 + {e^{2\gamma }}} \right)}^2}}}\left[ { - 2 - {e^\gamma } - 2{e^{2\gamma }} + {e^{5\gamma }} + 2{e^\gamma }{{\left( {1 + {e^\gamma }} \right)}^2}\left( {x + \ln \left( {\coth \left( {\gamma /2} \right)} \right)} \right)} \right],  \nonumber  \\
\end{eqnarray}
\end{widetext} 
where ${\rm{Li}}_2(z)$ is the polylogarithm function. In the limit of $r \to 0$, we have
\begin{eqnarray}\label{eq4.8}
\langle t_m \rangle=\frac{{{\pi ^2}{x_0}}}{{16\sqrt {Dr} }} \sim {r^{ - 1/2}}.
\end{eqnarray}
As expected, $\langle t_m \rangle$ diverges in the absence of resetting, and possesses the same asymptotic behavior as the mean first passage time $\langle t_f \rangle$ with $r\to 0$ \cite{evans2011diffusion,evans2020stochastic}. 

In Fig.\ref{fig4}, we show $\langle t_m \rangle$ as a function of $r$ for three different values of $x_0$, but for a fixed $D=1/2$. Interestingly, $\langle t_m \rangle$ show a nonmonotonic dependence on $r$ or equivalently on $\gamma$ for the fixed $x_0$ and $D$. There exist an optimal  $r$ at which $\langle t_m \rangle$ attains its unique minimum. Aslo, we have performed simulations (shown by symbols in Fig.\ref{fig4}), which are in excellent agreement with our theory (shown by lines in Fig.\ref{fig4}).  Taking the derivative of $\langle t_m \rangle$ with respect to $r$, and then letting the derivative equal to zero, we obtain the optimal $r$, determined by the equation $2 F_2(\gamma^{*})=\gamma^{*} F'_2(\gamma^{*}) $. The equation can be solved by numeric to yield $\gamma^{*}\approx 1.64831$, or equivalently, the optimal resetting rate
\begin{eqnarray}\label{eq4.9}
r^{*} \approx 2.71691 D/x_0^2.
\end{eqnarray}
We note that the mean first passage time through the origin shows a nonmonotonic change with $r$ as well, and has a minimum at an optimal $r$, which equals approximately to $2.53964 D/x_0^2$ \cite{evans2011diffusion,evans2020stochastic},  different from the value given in Eq.(\ref{eq4.9}). 

\section{Conclusions}
In conclusion, we have studied the extremal statistics of the maximal displacement $M$ of the RBM starting from a positive position $x_0$ and the time $t_m$ at which the maximum is reached before the first passage time through the origin. In the first part of this paper, we compute the exit probability of the RBM in an interval $\left[0, M \right] $ from the origin (Eq.(\ref{eq1.3})), and thus obtain the marginal distribution $P_r(M|x_0)$ (Eq.(\ref{eq1.5})). In particular, the expectation of $M$, $\langle M \rangle$, is obtained explicitly (Eq.(\ref{eq1.7})). We find that $\langle M \rangle$ decreases monotonically with the resetting rate $r$, and converges to its asymptotic value $2 x_0$ as $r \to \infty$, but diverges with the negative logarithm of $r$ as $r \to 0$. In the second part of this paper, we compute the survival probability and propagator of the RBM in an inverval $\left[0, M \right] $ with absorbing boundaries at both ends. However, these quantities are obtained explicitly only in the Laplace space. Subsequently, we obtain in the Laplace space the joint distribution of $M$ and $t_m$ based on a path decomposition technique for Markov processes. Fortunately, the expectation $\langle t_m \rangle$ of $t_m$ is obtained explicitly (Eq.(\ref{eq4.7})).  $\langle t_m \rangle$ diverges as $r \to 0$ with $\langle t_m \rangle \sim r^{-1/2}$, as the diffusing particle in the absence of resetting can meander indefinitely in the positive direction without reaching its maximum displacement. Also $\langle t_m \rangle$ diverges as $r \to \infty$, because that the diffusing particle has less time between resets to reach the origin as the resetting rate increases. In between these two divergences there is an optimal resetting rate (Eq.(\ref{eq4.9})) at which $\langle t_m \rangle$ attains its unique minimum. Such a phenomenon is reminiscent of mean first passage time $\langle t_f \rangle$ of the RBM, wherein $\langle t_f \rangle$ also possesses a single minimum at some resetting rate. We should emphasize that the two optimal resetting rates are quantitatively different. The optimal resetting for $\langle t_m \rangle$ is slightly larger than that for  $\langle t_f \rangle$. Therefore, our study provides an additional example regarding the nontrivial effects of stochastic resetting. In the future, it would be interesting to investigate the extremal statistics of other types of Brownian motions under resetting before their first passage to an absorbing boundary, such as active Brownian motions \cite{scacchi2018mean,kumar2020active} and run-and-tumble motions \cite{evans2018run,santra2020run,bressloff2020occupation,singh2022extremal}.

\begin{acknowledgments}
This work was supported by the National Natural Science Foundation of China (11875069, 61973001) and the Key Scientific Research Fund
of Anhui Provincial Education Department.
\end{acknowledgments}


\end{document}